\documentclass[journal]{IEEEtran}

\usepackage{xcolor,soul,framed} 

\colorlet{shadecolor}{yellow}
\usepackage[pdftex]{graphicx}
\graphicspath{{../pdf/}{../jpeg/}}
\DeclareGraphicsExtensions{.pdf,.jpeg,.png}

\usepackage[cmex10]{amsmath}
\usepackage{array}
\usepackage{mdwmath}
\usepackage{mdwtab}
\usepackage{eqparbox}
\usepackage{url}
\usepackage{newtxmath}
\usepackage{cite}
\usepackage{graphicx}
\usepackage{algorithm}
\usepackage{algpseudocode}
\usepackage{amsmath}
\usepackage{booktabs}
\usepackage{cite}
\usepackage{dsfont}   

\usepackage{amsmath,amsfonts}
\usepackage{algorithm}
\usepackage{algpseudocode}
\usepackage{array}

\usepackage{caption}
\usepackage{graphicx}
\usepackage{subcaption}        
\usepackage{stfloats}          

\begin{document}
\bstctlcite{IEEEexample:BSTcontrol}
    \title{A Novel Partitioning Scheme for RIS Identification and Beamforming}



\author{Yarkın~Gevez,~\IEEEmembership{Graduate Student Member,~IEEE}, Aymen Khaleel,~\IEEEmembership{Member,~IEEE}, and Ertugrul Basar,~\IEEEmembership{Fellow,~IEEE}
\thanks{
During this work, Yarkın Gevez, Aymen Khaleel, and Ertugrul Basar were with the Communications Research and Innovation Laboratory
(CoreLab), Department of Electrical and Electronics Engineering, Ko\c{c}
University, 34450 Sariyer, Istanbul.

Currently, Aymen Khaleel is with the Faculty of Electrical Engineering and Information Technology, Ruhr-University Bochum, Bochum 44801, Germany. (e-mail: aymen.khaleel@rub.de). Ertugrul Basar is with the Department of Electrical Engineering, Tampere University, 33720 Tampere, Finland (email: ertugrul.basar@tuni.fi).

}
}

\IEEEaftertitletext{\vspace{-2.5\baselineskip}}
\maketitle

\markboth{IEEE Wireless Communications Letters, VOL.~.., NO.~.., ..~2025
}{Roberg \MakeLowercase{\textit{et al.}}: High-Efficiency Diode and Transistor Rectifiers}



\begin{abstract}
This letter introduces a novel partitioning scheme for reconfigurable intelligent surfaces (RISs) that simultaneously consider RIS identification and beamforming. The proposed scheme dynamicly and efficiently allocates RIS elements between identification and beamforming users, considering the different performance metrics associated with each of them. By employing a dynamic partitioning algorithm that efficiently manage the RIS resources (elements), the scheme significantly enhances the signal-to-noise ratio (SNR) while maintaining reliable identification performance. Finally, theoretical analysis and computer simulations are provided to demonstrate the validity of the proposed scheme.

\end{abstract}



\vspace{-0.45cm}
\section{Introduction}
\vspace{-0.15cm}


Reconfigurable intelligent surfaces (RISs) are increasingly being recognized as a transformative technology for the physical layer in sixth-generation (6G) wireless communications. These surfaces consist of metamaterial elements that can be dynamically controlled, offering the potential to greatly enhance wireless systems by providing unprecedented manipulation of electromagnetic waves \cite{WirelessComm_Through_RIS}. By adjusting the phase, amplitude, and polarization of incoming signals, RISs can strengthen communication links, improve energy efficiency, and reduce interference \cite{RIS_for_6G}. The broad applicability of RISs has been explored in various wireless communication scenarios, such as non-orthogonal multiple access (NOMA) \cite{RIS-NOMA}, multiple-input multiple-output (MIMO) systems \cite{khaleel2020reconfigurable}, and index modulation techniques \cite{RIS-IM}, among others.

Beyond their applications in information communication systems, RISs have also been utilized in localization technologies. For instance, radio-frequency sensing systems that incorporate RISs have been developed for human posture recognition, emphasizing the importance of optimal RIS configuration to minimize recognition errors \cite{hu2020RIS_Sensing}. Hybrid RISs, which combine active reception elements for sensing and processing, have shown to enhance channel estimation efficiency by requiring significantly fewer pilot signals than fully reflective RISs \cite{Zhang_Hybrid_RIS}. Additionally, RIS-assisted integrated sensing and communication systems have been investigated in multi-user scenarios. In these systems, joint optimization of transmit beamforming and RIS settings is performed to strike a balance between communication performance and target detection capabilities \cite{Liu_ISAC}.


Recent papers \cite{basar2022reconfigurable}-\cite{cui2019secure} provide a comprehensive treatment of RIS hardware architectures, control protocols, and signal-processing techniques.  Although these works outline numerous communication enhancements enabled by RISs, they commonly presume that user devices already know which RISs are present and can uniquely identify them—a practical gap that our paper aims to close.

Recently, the authors in \cite{khaleel2024simple} studied the RIS detection and identification (RIS-ID) problem and introduced a simple scheme to address it. Building upon this foundation, our work proposes a novel partitioning scheme that jointly considers identification and beamforming within a single adaptive framework. By dynamically allocating RIS elements into two functional subsets, the proposed scheme balances detection probability for identification users and SNR performance for beamforming users, thereby enhancing overall system performance.

The remainder of this letter is structured as follows: In Section II, we introduce the system model, detailing the RIS configuration and the approach for simultaneous beamforming and identification in a two-user scenario. Section III presents the theoretical analysis, where we derive the outage probabilities for both unsorted and sorted RIS element allocation schemes. Finally, Section IV provides simulation results that validate the performance of the proposed scheme under various transmit power conditions.

\vspace{-3mm}
 \section{System Model}

We consider a single-cell wireless network in which a single-antenna transmitter ($T_x$) communicates with user equipments (UEs) via an RIS deployed to enhance wireless communication performance. RIS is uniquely identifiable through its predefined phase shift reflection pattern (PSRP) \cite{khaleel2024simple}. For clarity and tractability, this work focuses on a single-RIS configuration, although the proposed framework can be readily extended to multi-RIS scenarios.

\begin{figure}
  \begin{center}
  \includegraphics[width=3.1in]{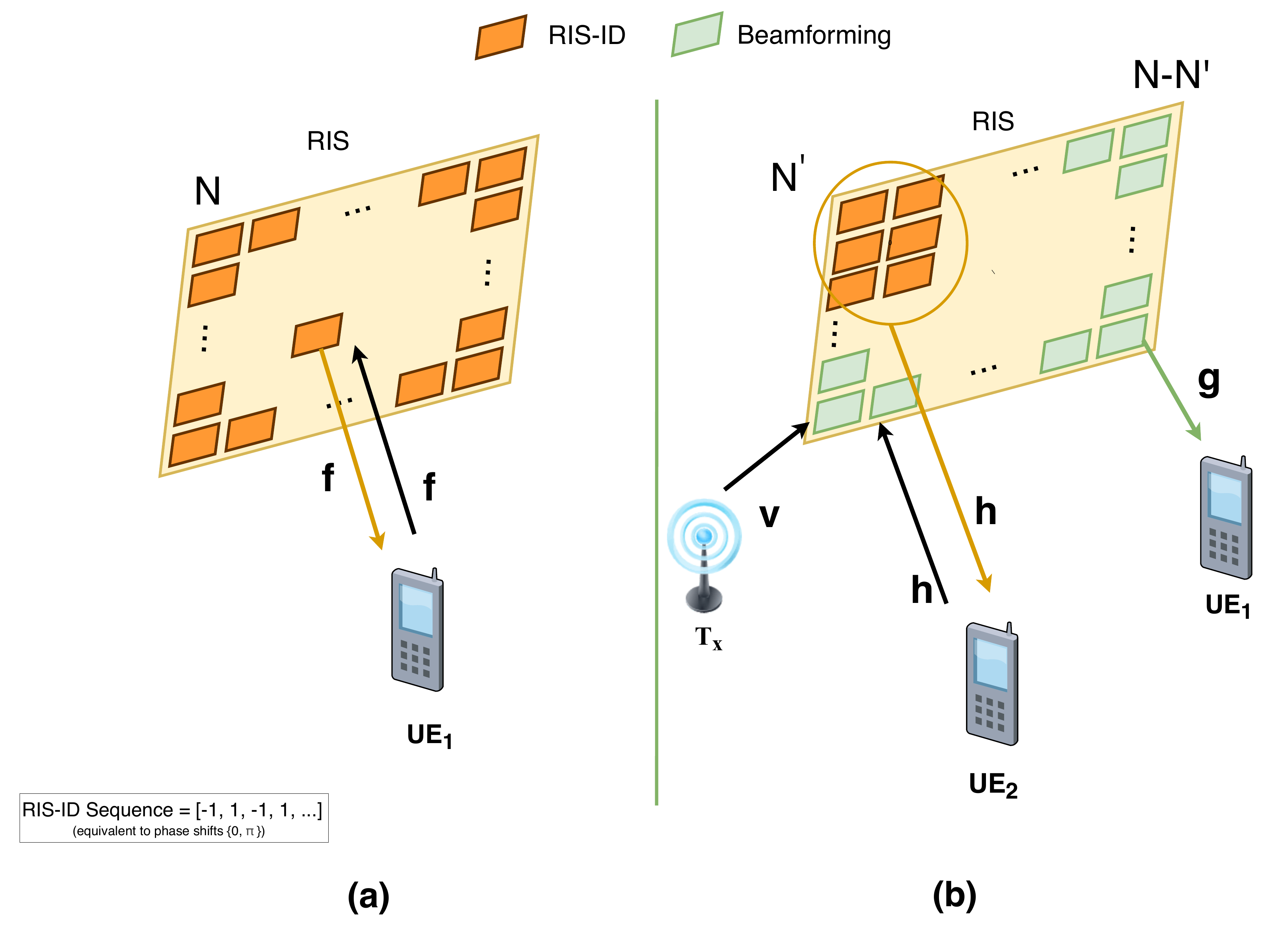}\\
  \caption{RIS partitioning scheme for two-user case:  
    (a) \emph{RIS-ID} phase  
    (b) \emph{Hybrid phase}: partitioned into RIS-ID and RIS-BF.}
    \label{scheme}
  \end{center}
\end{figure}

In the initial state, all $N$ elements operate in identification mode, reflecting a predefined spatial reflection pattern (PSRP) with fixed phase shifts of $0$ and $\pi$ to enable nearby UEs to detect the RIS.
 Once a UE successfully identifies the RIS, it begins communication through the RIS-assisted link, the RIS transitions to a hybrid operational mode in which its elements are dynamically partitioned between identification and beamforming tasks according to the proposed channel-aware allocation algorithm. This operational design ensures that the RIS continues to serve connected users while remaining discoverable to additional users in the environment.
As illustrated in Fig.~\ref{scheme} for the hybrid phase, the RIS serves two user equipments UEs: $UE_1$ (beamforming user) and $UE_2$ (identification user). The RIS comprises $N$ passive reflecting elements, which are dynamically divided into two disjoint subsets: the beamforming set (BFset, $\mathcal{BF}$) and the identification set (IDset, $\mathcal{ID}$). Specifically, $N - N'$ elements are assigned to $\mathcal{BF}$ to enhance the signal-to-interference-plus-noise ratio (SINR) for $UE_1$ via passive beamforming, while the remaining $N'$ elements are allocated to $\mathcal{ID}$ to enable RIS identification for $UE_2$.
\vspace{-2mm}

\vspace{-2mm}
\vspace{-0.1mm}

\subsection{Detection and Identification Phase}

Here, there is only $UE_1$ in front of the RIS and it tries to detect and identify nearby RISs as shown in Fig. \ref{scheme} (a). Initially, when the RIS has not yet been discovered by any user, all of its elements are allocated for RIS-ID purpose, detecting and identifying the RIS, $UE_{1}$ benefits from the entire RIS surface being dedicated to the identification process, ensuring an accurate detection before any partitioning or beamforming tasks are initiated. After the first UE connects to the RIS controller, some elements continue to serve the identification operation in the subsequent stages, as determined by the proposed dynamic allocation algorithm.
The received signal \( y_{m,1} \) at \( UE_{1} \), associated with the \( m \)-th symbol reflected by the RIS, can be written as
\begin{equation}
    y_{m,1} = x \left( \mathbf{f}_{m}^{T} \mathbf{\Theta}_m \mathbf{f}_{m} \right) + n_m,
\end{equation}

\noindent where $x = 1$ is the baseband sample of the unmodulated carrier signal transmitted from the full-duplex $\text{UE}_{1}$ \cite{keykhosravi2022ris}, and \( n_m \) is the additive white Gaussian noise (AWGN) component. Here, \( \mathbf{f}_m \in \mathbb{C}^{N \times 1} \) represents the complex baseband channel vector between the RIS and UE$_1$ during the identification phase as can be seen in Fig. \ref{scheme} (a).Here, we refer the reader to the system model and performance analysis associated with this phase detailed in \cite{khaleel2024simple} \footnote{Note that the RIS-ID concept proposed in [11] has been practically validated and tested in a real-world experiment in \cite{vural2025practical}, showing the potential of this novel concept.}.
\vspace{-3mm}
\subsection{Beamforming and Identification Phase}
In this phase, as shown in Fig. \ref{scheme}(b) consider two UEs ($\text{UE}_1$ and $\text{UE}_2$) where  $UE_1$ has already detected, identified, and connected with the RIS controller module to use the RIS for beamforming its signal of interest as explained in the previous subsection. This beamforming capability is achieved by continuously adjusting the phase shifts of each RIS element within the range of $[0, 2\pi]$, enabling coherent signal reflection toward the intended user without requiring active RF components.
Next, the RIS must dynamically partition its elements to fulfill two simultaneous functions. A subset of elements is allocated to perform passive beamforming for the user already connected to the RIS controller ($UE_1$), while the remaining elements are reserved for RIS identification (RIS-ID) to assist other potential users ($UE_2$) in discovering the RIS.
Let \( m \) denote the index of the received symbol at the UE. The channels from the UE to the RIS and from the RIS to the UE are denoted by \( \mathbf{h}_{UR,m} \) and \( \mathbf{h}_{RU,m}\in \mathbb{C}^{N \times 1} \), respectively. In addition, the vectors \( \mathbf{v}_m \in \mathbb{C}^{N \times 1} \) and \( \mathbf{g}_m \in \mathbb{C}^{N \times 1} \) represent the channel from the RIS to the user and the channel from the BS to the RIS, respectively. 
Both $\mathbf{v}_m \sim \mathcal{CN}(\mathbf{0},P_t\beta_v\mathbf{R}), \forall m,$ and $\mathbf{g}_m \sim \mathcal{CN}(\mathbf{0},P_t\beta_g\mathbf{R}), \forall m,$ are modeled as Rayleigh fading channels,  
where $\mathcal{CN}(\mathbf{0}, \sigma^2)$ denotes the complex Gaussian distribution with zero mean and variance $\sigma^2$. Here, $P_t$ is the transmit power, $\beta_v$ and $\beta_g$ represent the path gains for the $\mathbf{v}_m$ and $\mathbf{g}_m$ channels respectively, $\mathbf{0}$ denotes the $N$-dimensional zero vector, and $\mathbf{R}\in \mathbb{C}^{N\times N}$ is the RIS spatial correlation matrix constructed as in~\cite{bjornson2020rayleigh}. The correlation between the $w$-th and $\tilde{w}$-th elements is given by $\mathrm{sinc}\!\left(\tfrac{2\|\mathbf{u}_w - \mathbf{u}_{\tilde{w}}\|}{\lambda}\right)$. Here, $\mathbf{u}_w$ is the position vector of the $w$-th element, determined by its horizontal and vertical indices and spacing and $\lambda$ is the signal wavelength.

Here, \(\mathbb{C}\) denotes the set of complex numbers. It is assumed that the RIS–UE$_2$ link exhibits channel reciprocity under a time-division duplex (TDD) framework, where the uplink and downlink channels are identical and operate at the same frequency during the identification process. Under the channel reciprocity assumption, we have $ \mathbf{h}_{UR,m} = \mathbf{h}_{RU,m} = \mathbf{h}_{m}\sim\mathcal{CN}(\mathbf{0},P_t\beta\mathbf{R}), \forall m $.  The RIS–UE channel is modeled as Rayleigh fading. Also, $\mathbf{\Theta}$ denotes the RIS phase shift matrix, where $\mathbf{\Theta}_m=\mathrm{diag}(e^{j\theta_m^{1}}, \ldots, e^{j\theta_m^N})$ matrix can be partitioned into two subsets representing the phase shift matrices associated with the IDset and BFset elements. This allocation is based on instantaneous channel gain metrics, where elements with stronger gains toward the connected UE are selected for beamforming, while the rest continue to assist potential users through identification, as detailed in the following sub-section. The ID-set does not apply any intelligent beamforming; only the elements assigned to the BF-set are phase-optimized to participate in RIS-assisted communication. Then, the received signal at $UE_2$ for the \textit{m}-th reflected symbol, denoted by $y_{m,2}$, can be given as 

\vspace{-5mm}
\begin{align}
\small
y_{m,2} 
&= x \left[(\mathbf{h}_m)^{T} \mathbf{\Theta}_m \mathbf{h}_m\right] 
 + q_m \left[(\mathbf{h}_m)^{T} \mathbf{\Theta}_m \mathbf{v}_m\right] + n_{m,2} \nonumber \\[2pt]
&= \underbrace{x \sum_{i \in \mathcal{ID}} h^{2}_{m,i} e^{j\phi_{m,i}}}_{\text{signal by IDset}}
 + \underbrace{x \sum_{i \in \mathcal{BF}} h^{2}_{m,i} e^{j\theta_{m,i}}}_{\text{signal by BFset }} \nonumber \\[2pt]
&\quad + \underbrace{q_m \sum_{i \in \mathcal{ID}} v_{m,i} e^{j\phi_{m,i}} h_{m,i}}_{\text{interference by IDset}}
 + \underbrace{q_m \sum_{i \in \mathcal{BF}} v_{m,i} e^{j\theta_{m,i}} h_{m,i}}_{\text{ interference by BFset}} 
 + n_{m,2},
\label{eqRecUE2a}
\end{align}

and can be expressed in a more compact form as follows

\begin{align}
y_{m,2} 
&\overset{\text{(I)}}{=} x \tilde{h}^{\text{ID}}_{2} 
 + x \tilde{h}^{\text{BF}}_{2} 
 + q_m \tilde{v}^{\text{ID}}_{2} 
 + q_m \tilde{v}^{\text{BF}}_{2} 
 + n_{m,2}.
\label{eqRecUE2}
\end{align}

\noindent where $N'$ is the number of reflecting elements, dedicated for the RIS-ID process. Here, $q_m$ denotes the modulated data symbol transmitted by the $T_x$ at the $m$-th interval as part of the RIS-assisted communication process through $UE_1$.

In the simplified form represented at step I of (\ref{eqRecUE2}), we have $ \tilde{h}^{\text{ID}}_{2} =\sum_{i\in\mathcal{ID}}^{N'} h^2_{m,i} e^{j\phi_{m,i}} $ to represent the aggregated contribution from the RIS identification elements. The term \( \tilde{v}^{\text{ID}}_{2} = \sum_{i\in\mathcal{ID}}^{N'} v_{m,i} e^{j\phi_{m,i}} h_{m,i} \) captures the interference contribution from the base station via the IDset. Similarly, we set \( \tilde{h}^{\text{BF}}_{2} = \sum_{i\in\mathcal{BF}}^{N-N'} h^2_{m,i} e^{j\theta_{m,i}} \) to represent the self-interference caused by the RIS beamforming elements, while \( \tilde{v}^{\text{BF}}_{2} = \sum_{i\in\mathcal{BF}}^{N-N'} v_{m,i} e^{j\theta_{m,i}} h_{m,i} \) includes the interference contribution from the base station via the BFset.

In light of (\ref{eqRecUE2}), the received signal $y_{m,1}$ for $UE_1$ is defined as:
\vspace{-3mm}
\begin{align}
y_{m,1} &= q_m \left( \tilde{v}^{\text{BF}}_1 + \tilde{v}^{\text{ID}}_1\right) + x \left( \tilde{h}^{\text{BF}}_1 + \tilde{h}^{\text{ID}}_1 \right) + n_{m,1},
\label{eq:y_corrected_single_RIS_with_theta_UE1}
\end{align}

\noindent where, we have \( \tilde{h}^{\text{ID}}_{1} = \sum_{i\in\mathcal{ID}}^{N'} h_{m,i} e^{j\phi_{m,i}} g_{m,i} \), \( \tilde{v}^{\text{ID}}_{1} = \sum_{i\in\mathcal{ID}}^{N'} v_{m,i} e^{j\phi_{m,i}} g_{m,i} \), \( \tilde{h}^{\text{BF}}_{1} = \sum_{i\in\mathcal{BF}}^{N-N'} h_{m,i} e^{j\theta_{m,i}} g_{m,i} \), and \( \tilde{v}^{\text{BF}}_{1} = \sum_{i\in\mathcal{BF}}^{N-N'} v_{m,i} e^{j\theta_{m,i}} g_{m,i} \).

\vspace{-3mm}
 \subsection{Elements Partitioning}

 The RIS divides its elements between IDSet and BFSet based on the CSI of $UE_1$, as follows. The elements with the highest channel gain are allocated for passive beamforming, while the others are left for identification. This is because, for passive beamforming, having high channel gain is directly proportional to the obtained beamforming gain, while for RIS-ID, channel gain does not directly affect the miss and false detection probabilities. Therefore, the SINR at $UE_1$ can be boosted without effectively degrading the RIS-ID performance at $UE_2$. 
 



The partitioning process  starts by calculating the channel gains for each RIS element using the channel coefficients $\mathbf{v}$ and $\mathbf{g}$. These gains are then sorted in descending order to prioritize elements with the highest gains. The $N/2$ elements with the highest gains are allocated to the BFset, enhancing the received signal quality at $UE_1$. The remaining elements are assigned to the IDset, ensuring the RIS continues to be detectable by other UEs ($UE_2$). Accordingly, IDset elements are adjusted for RIS-ID \cite{khaleel2024simple} while BFset elements are adjusted for passive  beamforming. In what follows, we measure the performance of elements sorting algorithm by deriving the outage probability at $UE_1$ as the user with the highest priority. The dominant computational step in the algorithm is the sorting of RIS elements based on their channel gain. The computational complexity of process is therefore $\mathcal{O}(N \log N)$, which remains efficient even for reasonably large RIS sizes.

\section{Theoretical Analysis of Outage Probability}
The outage probability is characterized as the probability that the SINR is lower than a predefined threshold \( r \):

\begin{equation}
P_{\text{out}} = \Pr\left( \text{SINR} < r\right).
\end{equation}

From (3), the SINR at \( UE_1 \) is expressed as:

\begin{equation}
\text{SINR}_{UE_1} = \frac{\left| \tilde{v}^{\text{BF}}_{1} \right|^2}{\left| \tilde{v}^{\text{ID}}_{1} + \tilde{h}^{\text{BF}}_{1} + \tilde{h}^{\text{ID}}_{1} \right|^2 + \frac{\sigma^2}{P_t}}.
\end{equation}

In what follows, we first derive the outage probability when the elements are randomly partitioned between the IDset and BFset, and then derive it for the ordered case according to proposed elements partitioning procedure. To the sake of simplicity, we substitute the SINR expression as

\begin{equation}
\small
A = \tilde{v}^{\text{BF}}_{1}, \quad B = \tilde{v}^{\text{ID}}_{1} + \tilde{h}^{\text{BF}}_{1} + \tilde{h}^{\text{ID}}_{1}.
\end{equation}



The term \(A = \tilde{v}_1^{\mathrm{BF}}\) corresponds to the effective combined signal at the receiver from the beamforming RIS elements. According to Appendix~A, \(\tilde{v}_1^{\mathrm{BF}}\) is a summation of i.i.d. products of Rayleigh fading coefficients, which leads to a non-zero mean complex Gaussian distribution due to the law of large numbers. Therefore, the term \(|A|^2\), being the squared magnitude of a non-zero mean complex Gaussian variable, follows a non-central chi-square distribution with one degree of freedom.
On the other hand, \(B = \tilde{v}_1^{\mathrm{ID}} + \tilde{h}_1^{\mathrm{BF}} + \tilde{h}_1^{\mathrm{ID}}\) represents the aggregate interference at the receiver. As detailed in Appendix~B, each component is a summation of zero-mean complex Gaussian terms, which yields a zero-mean complex Gaussian random variable overall. Hence, the magnitude squared \(|B|^2\) follows a central chi-square distribution with two degrees of freedom.

Assuming the transmit power \(P_t\) for both the $T_x$ and \( UE_1 \), (5) can be rewritten as

\begingroup
\small
\setlength{\abovedisplayskip}{2pt}
\setlength{\belowdisplayskip}{2pt}
\setlength{\abovedisplayshortskip}{1pt}
\setlength{\belowdisplayshortskip}{1pt}
\vspace{-2mm}
\begin{equation}
\text{SINR}_{UE_1} = \frac{|A|^2}{|B|^2 + \frac{\sigma^2}{P_t}}.
\end{equation}

Submitting (7) in (4), we get

\begin{equation}
P_{\text{out}} =  \Pr \left( |A|^2 < r \left( |B|^2 + \frac{\sigma^2}{P_t} \right) \right).
\end{equation}

Defining a new random variable \(\Upsilon = |A|^2 - r|B|^2\), (8) can be rewritten as

\begin{equation}
P_{\text{out}} = \Pr \left( \Upsilon < \frac{r\sigma^2}{P_t} \right),
\end{equation}

\noindent where, the characteristic function of \(\Upsilon\) is given as:

\begin{equation}
\Psi_{\Upsilon}(\omega) = \exp \left( \frac{j\omega \mu_A^2}{1 - 2j\omega \sigma_A^2} \right) \frac{1}{(1 - 2j\omega \sigma_A^2)^{0.5}(1 + 2j\omega r \sigma_B^2)}.
\end{equation}
\endgroup








This expression follows the statistical moment-generating function approach for SINR distributions, building on the Gil-Pelaez inversion theorem as detailed in \cite{mathai1992quadratic}.

\subsection{Outage Probability for Unsorted Element Allocation}
To find the cumulative distribution function (CDF) of \(\Upsilon\), we apply Gil-Pelaez’s inversion formula \cite{mathai1992quadratic}:
\vspace{-2mm}

\begin{equation}
  P(\Upsilon < y)
  = \frac{1}{2}
    - \frac{1}{\pi} \int_{0}^{\infty}
      \frac{\operatorname{Im}\left( e^{-j \omega y} \Psi_{\Upsilon}(\omega) \right)}{\omega}
      \, d\omega .
\end{equation}

\noindent where the integral is computed over the imaginary part of the product of the exponential term and the characteristic function. The upper limit for the integration is set appropriately to avoid numerical errors. Finally, we evaluate this expression for \( y = \frac{r \sigma^2}{P_t} \) to obtain the outage probability \( P_{\text{out}} \).

To analyze the outage probability, we derive the means and variances of the random variables \(A\) and \(B\). The mean and variance of \(A\) are derived based on the product of two independent and identically distributed (i.i.d) Rayleigh faded channels. The mean of \(A\) is \(\mu_A = (N - N') \sigma_{v,\text{BF}} \sigma_{g,\text{BF}} \frac{\pi}{4}\), and the variance of \(A\) is \(\sigma_A^2 = (N - N') \sigma_{v,\text{BF}}^2 \sigma_{g,\text{BF}}^2 \left(1 - \frac{\pi^2}{16}\right)\). For large values of \(N\), \(A\) can be approximated as a Gaussian random variable using the central limit theorem (CLT). Similarly, the mean and variance of \(B\) are determined by considering the components of \(B\), which are i.i.d. Rayleigh fading channel coefficients. The mean of \(B\) is zero, \(\mu_B = 0\). The variance of \(B\) is \(\sigma_B^2 = \sigma_{g,\text{ID}}^2 N' \sigma_{v,\text{ID}}^2 + \sigma_{g,\text{BF}}^2 (N - N') \sigma_{h,\text{BF}}^2 + \sigma_{g,\text{ID}}^2 N' \sigma_{h,\text{ID}}^2\). Like \(A\), for large \(N\), \(B\) can also be approximated as complex Gaussian random variable according to the CLT (see Appendix A and B for detailed derivation).
\vspace{-7mm}
\subsection{Outage Probability with Channel-Gain-Based Element Sorting}
In our proposed scheme, we enhance the channel gains by sorting the RIS elements based on their channel gain values and selecting the top BFset elements for beamforming. This sorting significantly improves the effective channel gain and, consequently, the SINR at the UE.

For the sorted case, the mean of the channel gain \(A\) is given by \(\mu_{A_{\text{sorted}}} = (N - N') \sqrt{\sigma_{v_{\text{BF}_{\text{sorted}}}}^2} \sqrt{\sigma_{g_{\text{BF}_{\text{sorted}}}}^2} \frac{\pi}{4}\), and the variance is \(\sigma^2_{A_{\text{sorted}}} = (N - N') \sigma_{v_{\text{BF}_{\text{sorted}}}}^2 \sigma_{g_{\text{BF}_{\text{sorted}}}}^2 \left(1 - \frac{\pi^2}{16}\right)\). 

Similarly, the variance of the noise term \(B\) in the sorted case is \(\sigma^2_{B_{\text{sorted}}} = N' \sigma_{v_{\text{ID}_{\text{sorted}}}}^2 \sigma_{g_{\text{ID}_{\text{sorted}}}}^2 + (N - N') \sigma_{h_{\text{BF}_{\text{sorted}}}}^2 \sigma_{g_{\text{BF}_{\text{sorted}}}}^2 + N' \sigma_{g_{\text{ID}_{\text{sorted}}}}^2 \sigma_{h_{\text{ID}_{\text{sorted}}}}^2\) (see Appendix B for detailed derivation).

\begin{figure*}[t]             
  \centering
  \begin{subfigure}[t]{0.315\textwidth}
    \centering
    \includegraphics[width=\linewidth]{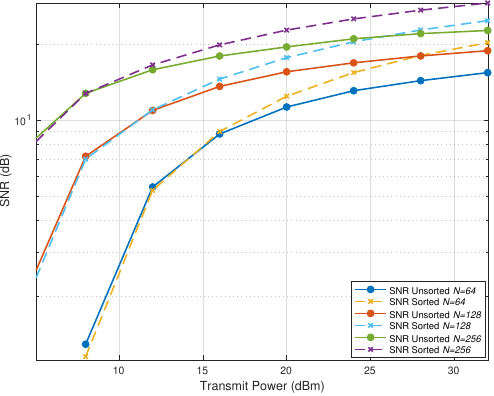}
    \caption{SNR comparison}
    \label{fig:snr_comparison}
  \end{subfigure}\hfill
  \begin{subfigure}[t]{0.323\textwidth}
    \centering
    \includegraphics[width=\linewidth]{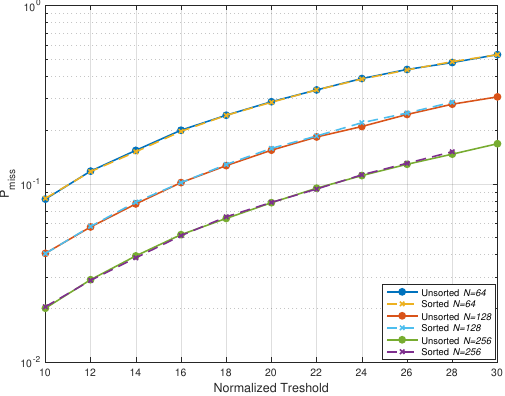}
    \caption{Miss detection probability ($P_{miss}$) for different thresholds}
    \label{fig:miss_detection}
  \end{subfigure}\hfill
  \begin{subfigure}[t]{0.31\textwidth}
    \centering
    \includegraphics[width=\linewidth]{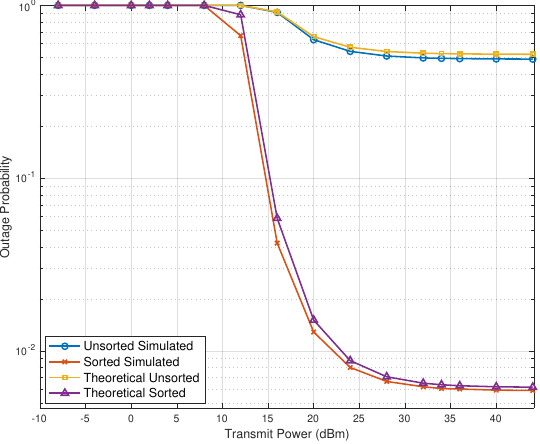}
    \caption{Outage probability}
    \label{fig:outage_probability}
  \end{subfigure}

  \caption{Performance metrics for sorted vs.\ unsorted RIS configurations.}
  \label{fig:three_side_by_side}
\end{figure*}


The CDF of the random variable for the sorted case is computed using Gil-Pelaez’s inversion formula  where \( \Psi_{\text{sorted}}(\omega) \) is the characteristic function of the sorted case, and the integral is computed over the imaginary part of the product of the exponential term and the characteristic function. The upper limit for the integration is set appropriately to avoid numerical errors. Finally, we evaluate this expression for \( y = \frac{r \sigma^2}{P_t} \) to obtain the outage probability \( P_{\text{out}_{\text{sorted}}} = P(Y_{\text{sorted}} < y) \) (see Appendix A and B for detailed derivation).



\vspace{-4mm}
\section{Simulation Results}
\vspace{-1.5mm}

In this section, we present computer simulation results for the proposed elements partitioning algorithm. The simulations were conducted over a transmit power range from $-10$ dBm to $45$ dBm was considered, with the RIS configurations involving a total of $N$ elements, partitioned into BFset for beamforming and IDset for identification. The wavelength $\lambda$ was calculated based on a carrier frequency of $1.8$ GHz. The path loss for the channels was computed using the standard formula, and the channels were modeled as Rayleigh fading. The noise power was assumed to be $-130$ dBm, which is a practical value for uplink communications. The simulations in Fig. 2(a) showed that the sorted RIS configuration consistently outperformed the unsorted configuration, particularly at higher transmit power levels.

 Specifically, as shown in Fig. 2(a), the sorted RIS partition consistently outperforms the unsorted configuration, especially at higher transmit power levels. This result underscores the effectiveness of the proposed algorithm in optimizing RIS 
 configuration to achieve better signal enhancement.


Figure~2(b) compares the miss–detection probability $P_{\text{miss}}$ of the baseline (unsorted) and proposed channel-gain–sorted RIS configurations. The derivation and analysis of $P_{\text{miss}}$ used in this comparison are provided in detail in~\cite{khaleel2024simple}. For all array sizes ($N\!=\!64,128,256$) and for the entire range of the normalized threshold $\bar r$, the two curves overlap, indicating that channel-gain sorting does \emph{not} impair identification accuracy.  Coupled with the SNR gains observed in Fig.~2(a), this confirms that the proposed partitioning strategy enhances communication performance while preserving reliable RIS identification.




Figure 2(c) compares the outage probability of the baseline (unsorted) RIS configuration with that of the proposed channel-gain-sorted configuration over a wide transmit-power range.  Both theoretical predictions and Monte-Carlo simulation results are plotted, and their close agreement validates the accuracy of the analytical model. The sorted configuration achieves up to an order-of-magnitude reduction in outage probability thanks to the SNR gain provided by allocating the strongest elements to the BFset.  As the transmit power exceeds roughly 25 dBm, the curves converge, indicating that the link becomes noise-limited rather than element-allocation-limited.  These results confirm that the proposed sorting algorithm can improve reliability without compromising the fidelity of the theoretical analysis.

In summary, the results confirm the effectiveness of the proposed joint beamforming and identification scheme, which integrates a dynamic partitioning algorithm and element sorting to optimize SNR performance. While the element sorting plays a key role in enhancing the SNR, the proposed scheme also maintains reliable identification accuracy. This balance between beamforming and identification demonstrates the practical utility of the system in dynamic wireless environments, all while preserving the expected SINR distribution.
\vspace{-6.7mm}

\vspace{-0.2cm}
\section{Conclusion}

This letter has presented a novel partitioning scheme for RISs that integrates user identification and beamforming in a dynamic and adaptive manner. The proposed scheme enhances SNR by efficiently sorting and allocating RIS elements based on channel gains. Importantly, this improvement in SNR is achieved without degrading the miss detection performance, ensuring reliable identification performance. Simulation results have validated the effectiveness of the proposed scheme, demonstrating its ability to optimize communication performance in varying channel conditions while maintaining robust identification accuracy. The strong alignment between theoretical and simulation results underscores the practical utility and robustness of this approach in dynamic wireless networks. Future work can explore extending this scheme to scenarios with multiple RISs or incorporating machine learning techniques for further optimization.

\vspace{-3mm}
\appendix
\subsection{Mean and Variance of \(A\)}
\label{Appendix B}

In this section, we derive the mean and variance of the random variable \(A\). The channels are assumed to be independent and identically distributed (i.i.d.) Rayleigh faded.

Let \(A\) be equal to $\tilde{v}^{\text{BF}}_{1}$ where \(\tilde{v}^{\text{BF}}_{1} = \sum_{i \in \mathcal{BF}}^{N-N'} v_{m,i,\text{BF}} e^{j \theta_{m,i}} g_{m,i,\text{BF}}\) is the product of two i.i.d Rayleigh distributed random variables, \(v_{m,i,\text{BF}}\) and \(g_{m,i,\text{BF}}\).

For Rayleigh distributed random variables \(v_{m,i,\text{BF}}\) and \(g_{m,i,\text{BF}}\) with parameters \(\sigma_{v,\text{BF}}\) and \(\sigma_{g,\text{BF}}\), respectively, the expected values are $\mathbb{E}[v_{m,i,\text{BF}}] = \sigma_{v,\text{BF}} \sqrt{\frac{\pi}{4}}$ and $\quad \mathbb{E}[g_{m,i,\text{BF}}] = \sigma_{g,\text{BF}} \sqrt{\frac{\pi}{4}}$.


Therefore, the mean of \(A\) is given by \(\mathbb{E}[A] = \mathbb{E}[\tilde{v}^{\text{BF}}_{1}] = \sigma_{v,\text{BF}} \sigma_{g,\text{BF}} \frac{\pi}{4}\). The variance of \(A\) can be calculated as \(\text{Var}[A] = \text{Var}[\tilde{v}^{\text{BF}}_{1}] = \mathbb{E}[|\tilde{v}^{\text{BF}}_{1}|^2] - (\mathbb{E}[\tilde{v}^{\text{BF}}_{1}])^2\). Since \(\tilde{v}^{\text{BF}}_{1} = v_{m,i,\text{BF}} g_{m,i,\text{BF}}\), with both \(v_{m,i,\text{BF}}\) and \(g_{m,i,\text{BF}}\) Rayleigh distributed, we have \(\mathbb{E}[|\tilde{v}^{\text{BF}}_{1}|^2] = \sigma_{v,\text{BF}}^2 \sigma_{g,\text{BF}}^2\). Therefore, \(\text{Var}[\tilde{v}^{\text{BF}}_{1}] = \sigma_{v,\text{BF}}^2 \sigma_{g,\text{BF}}^2 \left(1 - \frac{\pi^2}{16}\right)\), and the overall variance of \(A\) becomes \(\text{Var}[A] = (N - N') \cdot \sigma_{v,\text{BF}}^2 \sigma_{g,\text{BF}}^2 \left(1 - \frac{\pi^2}{16}\right)\).



According to the CLT, as the number of RIS elements \(N\) increases, the distribution of \(A\) approaches a Gaussian distribution. Given the calculated mean and variance, \(A\) can be approximated as
\vspace{-1mm}
{\small
\begin{equation}
A \sim \mathcal{N} \left((N - N')  \sigma_{v,\text{BF}} \sigma_{g,\text{BF}} \frac{\pi}{4}, \; (N - N') \sigma_{v,\text{BF}}^2 \sigma_{g,\text{BF}}^2 \left(1 - \frac{\pi^2}{16}\right) \right)
\end{equation}
}

\vspace{-2mm}
\subsection{Mean and Variance of \(B\)}
\label{AppendixC}

In this section, we derive the mean and variance of the random variable \(B\). The channels are assumed to be independent and identically distributed (i.i.d) Rayleigh faded.

Let \(B\) be defined as B =  $\tilde{v}^{\text{ID}}_{1} +  \left( \tilde{h}^{\text{BF}}_{1} + \tilde{h}^{\text{ID}}_{1} \right)$, where \(\tilde{v}^{\text{ID}}_{1}\), \(\tilde{h}^{\text{BF}}_{1}\), and \(\tilde{h}^{\text{ID}}_{1}\) are the products of two i.i.d Rayleigh distributed random variables.

Since there is no coherent phase alignment, the mean of \(B\) is zero as $ \mathbb{E}[B] = \mathbb{E} \left[ \tilde{v}^{\text{ID}}_{1} \right] + \mathbb{E} \left[ \left( \tilde{h}^{\text{BF}}_{1} + \tilde{h}^{\text{ID}}_{1} \right) \right] = 0. $

The variance of \(B\) is given by the sum of the variances of the individual components: \(\text{Var}[B] = \text{Var} \left( \tilde{v}^{\text{ID}}_{1} \right) + \text{Var} \left( \tilde{h}^{\text{BF}}_{1} \right) + \text{Var} \left( \tilde{h}^{\text{ID}}_{1} \right)\). Each \(\tilde{v}^{\text{ID}}_{1}\) follows a complex normal distribution \(\mathcal{CN}(0, \sigma_{v,\text{ID}}^2 \sigma_{g,\text{ID}}^2)\), resulting in a variance of \(N' \sigma_{v,\text{ID}}^2 \sigma_{g,\text{ID}}^2\). Similarly, \(\tilde{h}^{\text{BF}}_{1}\) follows \(\mathcal{CN}(0, \sigma_{h,\text{BF}}^2 \sigma_{g,\text{BF}}^2)\), with a variance of \((N - N') \sigma_{h,\text{BF}}^2 \sigma_{g,\text{BF}}^2\). Lastly, \(\tilde{h}^{\text{ID}}_{1}\) follows \(\mathcal{CN}(0, \sigma_{h,\text{ID}}^2 \sigma_{g,\text{ID}}^2)\), yielding a variance of \(N' \sigma_{h,\text{ID}}^2 \sigma_{g,\text{ID}}^2\). Summing these variances gives \(\text{Var}[B] = N' \sigma_{v,\text{ID}}^2 \sigma_{g,\text{ID}}^2 + (N - N') \sigma_{h,\text{BF}}^2 \sigma_{g,\text{BF}}^2 + N' \sigma_{h,\text{ID}}^2 \sigma_{g,\text{ID}}^2\).



According to the CLT, as the number of RIS elements \(N\) increases, the distribution of \(B\) approaches a Gaussian distribution. Given the calculated mean and variance, \(B\) can be approximated as
\vspace{-1mm}
{\small
\begin{equation}
B \sim \mathcal{N} \left( 0,  N' \sigma_{v,\text{ID}}^2 \sigma_{g,\text{ID}}^2 + (N - N') \sigma_{h,\text{BF}}^2 \sigma_{g,\text{BF}}^2 + N' \sigma_{h,\text{ID}}^2 \sigma_{g,\text{ID}}^2 \right)
\end{equation}
}

\vspace{-2mm}

\bibliographystyle{IEEEtran}
\bibliography{Bibliography}


\ifCLASSOPTIONcaptionsoff
  \newpage
\fi

\vfill


\end{document}